
\def\bigtype{\let\rm=cmr12  \let\it=cmti12 \let\bf=cmbx12
\let\sl=cmsl12
 \baselineskip=18pt }

\def\d#1/d#2{ {\partial #1\over\partial #2} }

\newcount\probno

\def\prob{\global\advance\probno by1 \the\probno }

\newcount\itemno
\def\item{  \global\advance\itemno by 1
(\romannumeral\the\itemno)
}

\newcount\probno
\def\prob{\global\advance\probno by1 \the\probno
\global\itemno=0}

\newcount\probsetno
\def\probset{\global\advance\probsetno by1
\centerline{Problem Set \the\probsetno } }

\newcount\topicno
\def\topic{\global\advance\topicno by1 \the\topicno }

\newcount\partno
\def\part{ PART\ \global\advance\partno by 1
\uppercase\expandafter{\romannumeral\the\partno}.
\global\probno=0\global\lectno=0
}

\newcount\lectno
\def\lect{ \centerline{LECT\ \global\advance\topicno by1
\the\lectno.} }

 \newcount\sectno
\def\sect #1 { {\it \global\advance\sectno by1 \the\sectno . #1} }

\def\pdr{\partial}

\def\ga{\gamma}

\def\eps{\epsilon}
\def\half{{1\over 2}}
\def\tr{\hbox{tr}}

\def\linebreak{\hfil\break}

\def\rochester{
 \centerline{\it Department of Physics and Astronomy}
 \centerline{\it University of Rochester}
 \centerline{\it Rochester, N.Y. 14627}
               }


\newcount\eqnumber
\def\beq{ \global\advance\eqnumber by 1 $$ }
\def\eeq{ \eqno(\the\eqnumber)$$ }
\def\n{\global\advance \eqnumber by 1\eqno(\the\eqnumber)}
\def\puteqno{
\global\advance \eqnumber by 1 (\the\eqnumber)}
\def\beqs{$$\eqalign}
\def\eeqs{$$}


\def\ifundefined#1{\expandafter\ifx\csname
#1\endcsname\relax}

\newcount\refno \refno=0  
\def\[#1]{
\ifundefined{#1}
\advance\refno by 1
\expandafter\edef\csname #1\endcsname{\the\refno}
\fi[\csname
#1\endcsname]}
\def\refis#1{\noindent\csname #1\endcsname. }

\def\label#1{
\ifundefined{#1}
\expandafter\edef\csname #1\endcsname{\the\eqnumber}
\else\message{label #1 already in use}
\fi{}}
\def\(#1){(\csname #1\endcsname)}
\def\eqn#1{(\csname #1\endcsname)}

\baselineskip=15pt
\parskip=10pt

\magnification=1200

\def\BEGINIGNORE#1ENDIGNORE{}

\baselineskip=20pt

\def\un#1{\underline{#1}}


\def\Gr{ \hbox{\rm Gr}} 

\magnification=1200
UR 1255

ER 40685-707

hept@xxx/9204075
\vskip.5in

\centerline{\bf The Effective Lagrangian of Three Dimensional} \centerline{\bf
Quantum Chromodynamics}
\vskip.5in
\centerline{G. Ferretti, S.G.Rajeev and Z. Yang}

\rochester
\vskip.5in
\centerline{\bf Abstract} We consider the low energy limit of three dimensional
Quantum Chromodynamics with an even number of flavors. We show that Parity is
not spontaneously broken, but  the global (flavor) symmetry is spontaneously
broken. The low energy effective lagrangian is a nonlinear sigma model on the
Grassmannian. Some  Chern--Simons  terms are necessary in the lagrangian to
realize the discrete symmetries  correctly. We consider also another
parametrization of the low energy sector which leads to a  three dimensional
analogue of the Wess--Zumino--Witten--Novikov model. Since three dimensional
QCD is believed to be a model for quantum anti--ferromagnetism, our effective
lagrangian can describe  their long wavelength excitations (spin waves).

\vfill\eject

\sect {Introduction}

In this and an accompanying  paper\[II] we will study the low energy behaviour
of Three Dimensional Quantum Chromodynamics  (3DQCD) by adapting  effective
lagrangian methods well known\[balreview]  in the theory of strong
interactions. The main motivation is to construct a lower dimensional
(therefore simpler) analogue of QCD where phenomena such as chiral symmetry
breaking can be studied. Furthermore three dimensional  QCD is of some
intrinsic interest   as a high temperature limit\footnote{$^*$}{  If we take
the high temperature limit of $\tr e^{-\beta H}$ in QCD, only the bosonic
(gluon) fields will contribute. However, the high temperature limit of $\tr
(-1)^Fe^{-\beta H}$ will have contributions from the quarks as well; the infra
red behaviour of this free energy can be studied by our effective lagrangians.}
of QCD. Another motivation  that three dimensional QCD it is believed to
describe quantum anti--ferromagnetism.

One point of view\[anderson],\[fradkin] is that a quantum Heisenberg model  on
the plane, of spin ${j}$ with $SU(2)$ symmetry is described by 3DQCD with $2j$
colors and two flavors. More generally, 3DQCD with  $SU(N_c)$ gauge symmetry
and $SU(N)$ flavor symmetry is related to a quantum anti--ferromagnet with
`spins' in a representation of $SU(N)$ with $N_c$ rows in the Young Tableau.
Such generalizations of the Heisenberg model have been studied in the condensed
matter literature \[affleck]. In another point of view\[wiegmann], an ordered
phase of the anti--ferromagnet in which the spins are periodic with a period
equal to a multiple of  the lattice spacing is considered. Then, it is possible
to reformulate the theory in terms of  a super--lattice; each point of the new
lattice represents a group of spins of the original lattice. Thus the field
variable on the new lattice will be a  multi--component  fermionic variable
that  satisfies certain bilinear  constraints.  The continuum limit of this can
be QCD. Of particular interest are anti--ferromagnetism in two space
dimensions,
as the  High $T_c$ superconductors are materials of this type. It is possible
that the charge carriers of these superconductors are solitons and that the
mechanism that binds them into Cooper pairs is an exchange of spin waves.
Effective lagrangians of the type we study provide a systematic framework in
which such questions can be answered. Such a close connection between
anti--ferromagnetism and gauge theories is an exciting prospect. We can apply
well--tested methods of the theory of strong interactions to make predictions
about anti--ferromagnetism. Conversely, new dynamical insights obtained from
studying anti--ferromagnetism can help us understand QCD.

In this paper, we will (except here in the introduction) use the particle
physics terminology. For clarity, it is important not to mix metaphors, and we
will stick to the one we feel more comfortable with.  For example, the
particles we call pions must be the spin waves of the anti--ferromagnet.
Electric charge of the anti--ferromagnetic system is to be identified with
baryon number in our language. Thus the spin waves  are electrically neutral.
The defects in the anti--ferromagnet caused by doping them would then
correspond to baryons (they are studied in an accompanying paper). Thus the
charge carriers  are the baryons. We discover that there is a mechanism (pion
exchange) by which  these charge carriers can form bound states of rather small
binding energy. Perhaps this can provide a mechanism for the formation of
Cooper pairs in the High $T_c$ superconductors, which have an
anti--ferromagnetic phase \[andersen].

The basic result of our paper is that 3DQCD has a parity preserving phase in
which chiral symmetry is spontaneously broken. In this phase, the low energy
behaviour of the  theory is very analogous to QCD. There are low mass
pseudo--scalars (`pions'), somewhat more massive vector mesons ($\rho$
mesons) and fermionic solitons of large mass (`baryons'). Although there could
be anyons even in a parity invariant theory, we find that do not arise in our
version of 3DQCD. Thus three dimensional QCD is a much better analogue for
strong interactions than two dimensional QCD, for example. There is no
spontaneous breaking of chiral symmetry in two dimensions; although a theory of
mesons and baryons as their solitons can be constructed \[witten]
\[forthcoming], in the ${1\over N_c}$ expansion, there is no systematic low
energy approximation.

We will first show that parity is not spontaneously broken in one version of
3DQCD. This is the theory with gauge group $SU(N_c)$ and an even number ($2n$)
of flavors: \beq L=-{1\over \alpha_s}\tr F^{\mu\nu}F_{\mu\nu} +\bar
q^i\ga\cdot\nabla q_i+m\left(\sum_{i>0} \bar q^i q_i-\sum_{i<0} \bar q^i
q_i\right). \eeq  The quark masses have been assumed to come in pairs of equal
magnitude but opposite signs, so that parity is a symmetry of the lagrangian.
The proof is an  adaptation  of the Vafa--Witten argument in four
dimensions\[vafawitten].

Then, we will argue that the  global symmetry $SU(2n)$ (analogous to chiral
symmetry),  is spontaneously broken to $S(U(n)\times U(n))$. We will then
construct a low energy effective lagrangian for this phase of 3DQCD. This will
a nonlinear sigma model on the Grassmannian $Gr_n=SU(2n)/S(U(n)\times U(n))$.
However, it will be necessary to add some Chern--Simons terms to this action so
that its doesn't have a discrete symmetry that 3DQCD does not
have\[wittcurralg]. The small oscillations of this  effective lagrangian
describes pseudo--scalars of small mass as well as vectors mesons of mass of
order $N_c^0$ in the ${1\over N_c} $ expansion. Some of the issues in this and
the accompanying paper \[II] are also discussed independently in \[lee].

\sect { Three Dimensional QCD and its Flavor Symmetries}

We will study the theory (Three Dimensional Quantum ChromoDynamics, 3DQCD),
defined by the action \beq S=\int \left[-{1\over \alpha_s}\tr
F^{\mu\nu}F_{\mu\nu} +\bar q^i\ga\cdot\nabla q_i+\sum_i m_i \bar q^i q_i\right]
d^3x. \eeq Here $q_i$ are two component spinors transforming under the
fundamental representation of the color group $SU(N_c)$; both spin  and color
indices are suppressed. The index $i=1\cdots N$ labels flavor; in the limit
where all the masses are equal, there is a global $U(N)$ symmetry associated to
flavor.  The $U(1)$ part of this symmetry, (with current  $J_{\mu}={1\over
N_c}\sum_i\bar q_i\ga_{\mu}q^i$ corresponding
 to baryon number), is a symmetry
for any choice of quark masses. There is no distinction between left and right
handed fermions in three dimensions, so the flavor symmetry does not have the
chiral form $U(N)_L\times U(N)_R$ familiar from four dimensions. Although
$U(N)$ symmetry allows equal  masses for all the quarks, assuming parity
symmetry  in addition will require that the quark masses are zero.
\footnote{*}{ Parity is the symmetry $P_1:A(x_1,x_2,t)\to A(-x_1,x_2,t),
q(x)\to q(-x_1,x_2,t),  \bar q(x)\to -\bar q(-x_1,x_2,t)$. We regard
$A=A_{\mu}dx^{\mu}$ as a one form.}  Thus the flavor symmetry in the limit of
massless quarks is $U(N)\times Z_2$. It is possible to have massive quarks in a
parity conserving theory if the quarks masses come in pairs with equal
magnitude but opposite sign.  If the number of flavors is even $N=2n$, the most
symmetric possibility with massive quarks is ($i=-n,\cdots,-1,1,\cdots n$),
\beq L=-{1\over \alpha_s}\tr F^{\mu\nu}F_{\mu\nu} +\bar q^i\ga\cdot\nabla
q_i+m\left(\sum_{i>0} \bar q^i q_i-\sum_{i<0} \bar q^i q_i\right). \eeq In this
case the global flavor symmetry is $S(U(n)\times U(n))\times Z_2$. This theory
is  invariant under a parity operation  ($P$) that is the product of the usual
parity operation ($P_1$) with the interchange $P_2:i\to -i$.

With an odd number of quarks, there is no way to make all of them   massive
without violating parity. This means that parity has a `global  anomaly', and
terms that violate parity can be induced due to radiative  corrections \[djt],
\[anr]. In particular,there is nothing to prevent a Chern--Simons term in the
color gauge field \[djt], \beq S_{\hbox{\rm C.S.}}=K\int \tr [AdA+{2\over 3}
A^3] \eeq This means that the theory exists in a phase where the color forces
are screened rather than confining.  Once such parity breaking is allowed, the
quarks can acquire a mass of order $g^2$, the gauge coupling constant. There
will be particles (quarks) carrying fractional baryon number and exotic
statistics ('non--abelian' fractional statistics) in the spectrum. The gluons
and quarks will have masses of the same order. This phase of 3DQCD is related
to the `generalized flux phase' of anti--ferromagnetism\[fradkin].

This phase is very different from the one we will study in detail, in which
parity is a symmetry. There are   light particles (`pions') that carry no
baryon number.  There are heavier  particles (`baryons') with  integer baryon
number, and fermi statistics. It is expected that all the particles that are
observable at infinity in a scattering experiment in QCD are color singlets;
there should be some phase in which this is true in three dimensions as well.
It should therefore be possible to construct a theory equivalent to the one
above, describing directly  these color singlet bound states. At present this
is possible only in two dimensional QCD. In three dimensional QCD, (as in four
dimensions), we should be able to  construct an effective lagrangian for the
low  mass particles of the theory, under reasonable assumptions. This is what
we will do in this paper.

The first issue is whether  the flavor symmetry $U(N)\times Z_2$ of the theory
is spontaneously broken in the limit of massless quarks.   If $N$ is even, an
analogue of the Vafa--Witten argument will show that parity is not
spontaneously broken.

Consider the expectation values of some operator $X$ involving only gluons that
changes sign under  parity (For example $X=\eps_{\mu\nu\rho}\tr
F_{\mu\sigma}\nabla_{\nu}F_{\rho\sigma}$). We will assume that the number of
flavors is even and the quarks have masses $m_i=-m_{-i}\neq 0$ that do not
break parity explicitly. Then the generating function for the expectation
values of $X$ is  \beq e^{-W[J]}=\int D[A] \det\left[i\ga\cdot\nabla +iM\right]
e^{\int [-{1\over \alpha_s}\tr F^{\mu\nu}F_{\mu\nu}+iJX]dx}. \eeq $M$ is the
diagonal mass matrix for the quarks. The fermion determinant  \beqs{
\det\left[i\ga\cdot\nabla +iM\right]&=
\prod_{i=1}^{n}\det[i\ga\cdot\nabla+im_i][i\ga\cdot\nabla-im_i]\cr
&=\prod_{i=1}^n\det[-(\ga\cdot\nabla)^2+m_i^2)]\cr }\eeqs is  a parity
invariant positive function of the gauge fields. It then follows that $W[J]$ is
an even convex function of $J$. The effective potential  $\Gamma[X]$ (the
Legendre transform of $W[J]$) is then also an even convex function. Hence the
minimum of $\Gamma[X]$ (the vacuum expectation value of $X$) is at $X=0$. Thus
parity will not be broken by the expectation value of any function of gauge
fields. The argument can be extended to functions of Fermi fields as well, as
remarked by Vafa and Witten. The reason this argument works is that there is a
parity invariant Infra Red regularization which makes  the fermion determinant
positive. The fermion determinant is not positive for odd $N$, and  parity
might  be broken by anomalies as well as spontaneously.

Now let us see if the continuous part of the flavor symmetry can be broken
spontaneously. Imagine that we have a small mass term $m \bar q \eps q$ added
to the lagrangian. ($\eps={\rm diag}(1,\dots,1,-1,\dots,-1)$) It is possible to
calculate the two point function of flavor currents \beq <j_{\mu}^a(k)
j_{\nu}^b(-k)>\sim {N_c \over 4\pi} \eps^{ab} \eps_{\mu\nu\rho}ik_{\rho} \;\;
\hbox{\rm for}\;\; k \to 0\eeq from a one loop diagram.  There is a theorem
stating that this long distance behaviour is in fact exact to all orders  in
perturbation theory \[noren]. Moreover, it remains non--zero even as $m\to 0$.
In this limit, the QCD lagrangian is invariant under $U(N)$, yet,  here is a
Green's function that is not invariant under $U(N)$. This is characteristic of
spontaneous symmetry breaking. (There is no anomaly in the flavor current  in
2+1 dimensions, so the symmetry is not explicitly broken by radiative
corrections. Thus, chiral symmetry breaking is a consequence of the {\it
absence} of flavor anomalies in three dimensions. In four dimensional QCD, the
't Hooft anomaly `matching conditions' \[thooft] are used usually to arrive at
the same result.)

If $U(N)\times Z_2$ is spontaneously broken, what is the unbroken subgroup?. We
will use an analogue of the Coleman--Witten argument \[colewitten] (based on
the large $N_c$ limit)  to show that, for even $N=2n$, $U(2n)$ breaks  to
$U(n)\times U(n)$. Thus the Goldstone bosons are described by a nonlinear sigma
model on the Grassmannian $Gr_n=U(2n)/U(n)\times U(n)$. Consider the effective
potential  $V(\Phi)$ of the order parameter $\Phi_i^j\sim \bar q_i q^j$ of
flavor symmetry breaking. It transforms under the adjoint representation of
$U(N)$ and $\Phi\to -\Phi$ under $P_1$. The effective potential can be expanded
in terms of the invariants of $U(N)$: \beq V(\Phi)=\sum_{n=1} v_n\tr
\Phi^n+\sum_{n=1,m=1}v_{n,m} \tr \Phi^n\tr \Phi^m+\cdots .\eeq In the large
$N_c$ limit, the terms involving products such as $\tr\Phi^n\tr \Phi^m$ are
suppressed: they involve more than one quark loop. To leading order in ${1\over
N_c}$, therefore, \beq V(\Phi)=\sum_{i=1}^{N} v(\phi_i) \eeq where $\phi_i$ are
the eigenvalues of the hermitian matrix $\Phi$, and $v(\phi)$ is some function
of a real variable. Since $\bar q q$ (and hence $\Phi$) changes sign under the
discrete symmetry $P_1$, this function must be even. The vacuum expectation
value of $\Phi$ will be determined by the minima of $v(\phi)$. If the minimum
is at a non--zero value of $\phi$, the $U(N)\times Z_2$ symmetry will be
broken. If $a$ is a minimum, so must be $-a$. It is reasonable to assume that
there are no other minima (i.e., that there is no accidental degeneracy). Then
the vacuum  expectation value of $\Phi$ will be of the form  \beq
<\Phi>=\hbox{\rm diag}(-a,-a,\cdots, a,\cdots a). \eeq This breaks $ U(N)\to
U(n)\times U(N-n)$. The value of $n$ (the number of negative eigenvalues)
cannot be determined at this order of the ${1\over N_c}$ expansion: we need a
term of the type $(\tr\;\Phi)^2$ in the effective potential. In order to
preserve parity, in some higher order of the ${1\over N_c}$ expansion, the
value $n={N\over 2}$ must be selected as the ground state.

\sect { Effective Lagrangian for 3DQCD}

The low energy effective lagrangian must describe  Goldstone bosons on the
Grassmannian, $\Gr_{n}={SU(2n)/G}$ where $G=S((U(n)\times U(n)))$.  (The
$U(1)$ of baryon number acts trivially  on the Grassmannian, so that
$U(2n)/U(n)\times U(n)$ is the same as $SU(2n)/S(U(n)\times U(n))$.)  This
theory  can be written in many ways, some of which we will discuss in a later
section. The description  that seems most convenient for our purposes is to use
the action, \beq S_1(\chi,A_{\mu})= {{F_\pi}\over 2}\int\tr
\nabla_{\mu}\chi^{\dag}\nabla^{\mu}\chi d^3x \eeq where $F_\pi$ is the analogue
of the pion decay constant. Here, $\chi\in SU(2n)$ and $A_\mu\in \un{G}$  are
the  basic field variables, and  \beq \nabla_{\mu} \chi=\pdr_{\mu}\chi+\chi
A_\mu. \eeq The gauge field is really a Lagrange multiplier that can be
eliminated by its equations of motion, \beq
A_{\mu}=-iP_{\un{G}}\left(\chi^{\dag}\pdr_{\mu}\chi\right) \eeq $P_{\un{G}}$
being the projection operator to the unbroken subalgebra $\un{G}$. The
Lagrangian has a gauge invariance under the right action of $G$ on $\chi$,  so
that the true degrees of freedom lie on the coset space $SU(n)/G$. There is,
also, a global symmetry under the left action of $U(2n)$.

There are also a pair of discrete global symmetries, $\sigma:\chi\to \chi
\pmatrix{0&1\cr 1&0\cr}$ and $P_2:\chi\to \pmatrix{0&1\cr 1&0\cr}\chi$, which
are related to parity (see below). Note that, unlike $P_2$, $\sigma$  does not
commute with the gauge invariance: under conjugation by $\sigma$ the gauge
transformation $\pmatrix{g_1&0\cr 0&g_2\cr}$ is mapped to  $\pmatrix{g_2&0\cr
0&g_1\cr}$. Thus  $\sigma$ is in the `normalizer' of the gauge group, which is
sufficient for its action on the right cosets to be well--defined.

Expanding $\chi$ around the trivial solution $\chi=1$ shows that there are a
set of $2n^2$ massless scalars, as needed.  The quantity
$\Phi=\chi\eps\chi^{\dag}$ transforms like the order parameter $\bar q \eps q$
of the symmetry breaking. \footnote{\dag}{ $\eps=\pmatrix{1&0\cr 0&-1\cr}$ is
an element of the Lie algebra $SU(N)$. The Grassmannian is  being  considered
as the co--adjoint orbit of $\eps$. A point on the Grassmannian can also be
parametrized by  a Hermitian matrix $\Phi$ satisfying $\tr \Phi=0, \Phi^2=1$.}

However, this lagrangian has too much symmetry.   The massless 3DQCD action is
invariant under the parity  $P_1$.  Under this, the order parameter transforms
as  \beq P_1:\Phi(x_1,x_2,t)\to -\Phi(-x_1,x_2,t) .\eeq In terms of $\chi$,
this can be thought of as the product of two transformations, \beq
P_0:\chi(x_1,x_2,t)\to \chi(-x_1,x_2,t)\quad  \sigma: \chi(x,t)\to
\chi(x,t)\pmatrix{0&1\cr 1&0\cr}. \eeq But for our action, $P_0$ and $\sigma$
separately are  symmetries.  The microscopic theory, 3DQCD, does not have such
a pair of symmetries, only their product is a symmetry. We need to break these
separate symmetries to their product, by adding some higher derivative terms to
the action.  The required terms are Chern--Simons terms for the gauge  field
\[djt], \[hagen]: \beq S_k={k\over 4\pi}\int \tr \left[ A_1dA_1+{2\over
3}A_1^3\right]-{k\over 4\pi}\int \tr \left[ A_2dA_2+{2\over 3}A_2^3\right].
\eeq Here, we have decomposed $A$  into components $A_1,A_2,A_3$ in the simple
factors $\un{SU(n)}\oplus \un{SU(n)} \oplus U(1)_{\eps}$ of $\un{G}$
respectively.  The operation $P_0:A_{1,2}(x,t)\to A_{1,2}(-x_1,x_2,t)$  will
change the sign of each Chern--Simons term, and $\sigma:A_1\leftrightarrow A_2$
will interchange them. So the theory is invariant under $P_1=P_0\sigma$. The
field $A_3$  changes sign under parity. It does not acquire a Chern--Simons
term as, it would  break  $P_1$.

The constant $k$ must be an integer for the action $e^{iS_k}$ to be gauge
invariant \[djt]. Furthermore, comparison  with 3DQCD will show that $k=N_c$.
(We can calculate the long range behaviour of the flavor current two point
function in QCD exactly. Comparing this to the prediction of the effective
lagrangian, we see that $k=N_c$.) Once these terms are added, the gauge fields
become propagating fields. In fact a linear analysis shows that the spectrum
consists of $2n^2$ massless scalars, plus, a set of $n^2$ spin one particles of
mass $\mu_A={2\pi F_\pi\over k}$. The positive helicity states come from $A_1$
and the negative helicity states from $A_2$.

If there is an explicit breaking of the $U(N)$ symmetry (by the addition of a
current quark mass) to $G$, there should be a corresponding mass term, \beq
S_m= {{F_\pi}\over 2} m_{\pi}^2\int \tr \eps\chi\eps\chi^{\dag} d^3x \eeq in
the lagrangian. This term is still invariant under the  {\it right} action of
the gauge group $G$. Furthermore, the global symmetry is now a   {\it left}
action by $G$. If we expand around  the trivial solution, we get a set of
pseudo-Goldstone bosons of mass $m_\pi$. If the explicit symmetry breaking is
small, $m_\pi<<F_{\pi}$, these bosons will be light (compared to the scale
that determines their self--interaction, $F_{\pi}$.) In addition, of course,
are the vector mesons of mass $2\pi{F_{\pi}\over k}$.

Thus we arrive at the following long distance  effective action for 3DQCD:
\beqs{ S&=S_1+S_k+S_m\cr &={{F_\pi}\over
2}\int\tr\nabla_{\mu}\chi^{\dag}\nabla^{\mu}\chi d^3x\cr &+{k\over 4\pi}\int
\tr \left[ A_1dA_1+{2\over 3}A_1^3\right]-{k\over 4\pi}\int \tr \left[
A_2dA_2+{2\over 3}A_2^3\right]\cr & + {{F_\pi}\over 2}m_{\pi}^2\int \tr
\eps\chi\eps\chi^{\dag} d^3x.\cr }\eeqs The spectrum of small oscillations of
this theory can be understood as mesons in the naive quark model. Let us call
the  quarks with a positive mass the `up' quarks and the others the `down'
quarks. In three dimensions, the Dirac equation predicts that quark  and
anti--quarks states will have positive helicity for positive mass and negative
helicity for negative mass. Thus the bound states of the type $\bar u u$, which
have positive helicity, can be identified with $A_1$, and the negative helicity
bound states $\bar d d$ with $A_2$. The bound states of type $\bar u d$ and
$\bar d u$ describe the spin zero states described by $\chi$. Thus our
effective lagrangian describes the spin zero and spin one mesons of 3DQCD. It
is possible to  find effective lagrangians also for the case where the symmetry
breaks in an uneven (parity  violating)  way $SU(N)\to S(U(N-n)\times U(n))$
by effective lagrangians. Some of these cases we will discuss in a later
section. They also have quark model interpretations.

In an accompanying paper, we will show that this effective  action has soliton
solutions. They will be shown to be the  baryons of 3DQCD.

\sect {Three dimensional analogue of the  Wess--Zumino--Witten--Novikov model}

We describe in this section some alternative approaches constructing the low
energy effective lagrangian. They are equivalent as far as the properties of
massless particles are concerned, but not for the massive particles. We find
that these effective lagrangians do not contain enough of the short distance
physics to produce baryons as solitons. They might be useful in describing
three dimensional critical phenomena.

In  a two dimensional space time $M$, the Wess--Zumino--Witten--Novikov model
has a field variable valued in a compact Lie Group $G$. The  action  of the
model\[witten2d],\[bhattraj] can be written as  \beq S=\half F_{\pi}\int_M \tr
g^{-1}dg * g^{-1} dg +{k\over 12 \pi^2} \int_{M_3} \tr \left(g^{-1}dg\right)^3.
\eeq Here the second term is an integral on a three dimensional manifold $M_3$
whose boundary is space time,$\pdr M_3=M$. The last term represents the third
co--homology of the configuration space (target space). This term breaks the
two separate discrete symmetries \beq P_1:g\to g^{\dag},\quad  x\to -x \eeq to
their product $P=P_1P_2$, so that the field variable describes a
pseudo--scalar. The current algebra of this model is a Kac--Moody algebra.
$F_\pi$ is a dimensionless variable so that theory is classically scale
invariant. This scale invariance is broken in the quantum theory, except for
two  values of $F_\pi$. At $F_\pi=\infty$ there is an Ultra--Violet stable
fixed point. Also, there is a non--trivial  IR  fixed point for this theory at
which it describes a conformal field theory.

It is of some interest to find a higher dimensional generalization of this
model. For example the current algebra of  such a model  could provide a
generalization for the Kac--Moody algebra.  To find an analogous  model in a
three dimensional space--time, we must have a target  space $M'$  with
$H^4(M')\neq 0$. One obvious candidate is $CP^{N-1}$, for which
$H^4(CP^{N-1})=Z$ for $N>2$. However, $\pi_4(CP^{N-1})=0$ so that this is not a
sufficiently close generalization.  A more appropriate generalization is  a
nonlinear model on the Grassmannian $Gr_{N,n}=SU(N)/S(U(N-n)\times U(n))$, with
$n>1$. Then, $\pi_4(Gr_{N,n})=Z$ and $H^4(Gr_{N,n})=Z\oplus Z$.

Each point on the Grassmannian $Gr_{N,n}$ describes an $n$ dimensional subspace
of $C^N$. By attaching to each point the corresponding $n$ dimensional vector
space, we get a natural complex vector bundle over $Gr_{N,n}$. The  Chern
classes of this vector bundle generate the cohomology ring of the Grassmannain.
In the limit $N\to \infty$, the topology of the Grassmannian simplifies
\[milnor], so that the homotopy groups can be
calculated:$\pi_{2l}(Gr_{\infty,n})=Z, \pi_{2l+1}(Gr_{\infty,n})=0$. It is
interesting that the dynamics of the theory also simplifies in this large $N$
limit, enough to make it exactly solvable. (This is a simple extension of the
work of ref. \[arefeva] on the $CP^{N-1}$ model.)

The Grassmannian can also be thought of as the set of all hermitian matrices of
square one, with the trace picking out the the particular connected component:
\beq Gr_{N,n}=\{\Phi|\Phi^{\dag}=\Phi,\,\,\Phi^2=1,\;\;\tr \;\Phi=2N-n\;\} \eeq
In our earlier parametrization as a coset, $\Phi=\chi\eps\chi^{\dag}$. The
generator of the even order cohomology is  \beq \omega_{2l}=\tr \Phi
\left(d\Phi\right)^{2l}. \eeq Then \beqs{       d\omega_{2l}&=\tr d\Phi
\left(d\Phi\right)^{2l}\cr &=\tr d\Phi \Phi^2 \left(d\Phi\right)^{2l}\cr
&=(-1)^{2l+1}\tr d\Phi \Phi \Phi \left(d\Phi\right)^{2l}\cr }\eeqs by moving
one of the $\Phi$'s through all the $d\Phi$'s; (recall that
$\Phi^2=1\Rightarrow \Phi d\Phi=-d\Phi \Phi$,)  . So, $\omega_{2l}$ is indeed
closed.  The two generators of $H^4$ are therefore $\omega_2^2$ and $\omega_4$.
($\omega_2$ is just the Kahler form).

The nonlinear sigma model on the Grassmannian, has action  \beq S_1=\half
F_{\pi}\int \tr d\Phi*d\Phi. \eeq This is invariant under the discrete
symmetries \beq \Phi(x)\to -\Phi(x) \;\;\hbox{\rm and }\;\; \Phi(x)\to
\Phi(-x_1,x_2,t) \eeq separately. As usual we can get a model invariant only
under the product $P_1$ of the two by  adding topological terms. Of the two
generators $\omega_2^2$ and $\omega_4$ of $H^4$, only $\omega_4$ is invariant
under $P_1$. Thus the three dimensional analogue of the
Wess--Zumino--Witten--Novikov  model has action  \beq S=\half F_{\pi}\int_M \tr
d\Phi*d\Phi+{k\over 64\pi}\int_{M_4} \tr \Phi (d\Phi)^4.        \eeq $M_4$ is a
four--manifold of which the space time manifold is the boundary. The numerical
constant in front of the second term  is chosen such that  the level number $k$
an integer.

The global symmetry under $SU(N)$ \beq \Phi\to g\Phi g^{\dag} \eeq  can be
gauged  by the Noether procedure  to arrive at the effective action, \beq
S=\half F_\pi  \int\tr (d\Phi+[V,\Phi])^{\dag}(d\Phi+[V,\Phi]) +
\Gamma[\Phi,V]. \eeq  Here, \beqs{  \Gamma[\phi,V]&= {k\over 64\pi} \int_{M_4}
\tr \Phi (d\Phi)^{4}\cr &+{k\over 8\pi} \int_{M} \tr \left[ V\Phi (d\Phi)^{2}+
dV \Phi V + dV V \Phi +(V\Phi)^{2}d\Phi + \Phi V^{3} +{1\over 3} (V \Phi)^{3}
\right].\cr }\eeqs It is possible to gauge the entire global symmetry $SU(N)$
unlike in the even dimensional WZWN models. Thus the topological term does not
represent an anomaly. Indeed, there is no anomaly in odd dimensional
space--time. The above effective action is gauge invariant only modulo some
surface terms, which means that it leads to the existence of some `edge
currents' at the surface.

Unlike the two dimensional WZWN model, this theory is not classically scale
invariant. ($F_\pi$ has dimensions of mass). Hence the loop expansion of this
theory is not renormalizable.  However, there is another expansion method that
can be used to define the quantum theory, the ${1\over N}$ expansion. In the
limit $N\to \infty$ keeping $n$ fixed, we have a sensible expansion. It is best
to rewrite this theory in yet another parametrization, \beq \Phi=2ZZ^{\dag}-1
\eeq where $Z$ is an $N\times n$ matrix with  \beq Z^{\dag}Z=1. \eeq The action
equivalent to the nonlinear sigma model on the Grassmannian is  \beq
S_1(Z,A)=\half \int \tr \nabla^{\mu}Z^{\dag}\nabla_{\mu}Z d^3x+ N\int \tr
\sigma(Z^{\dag}Z-{1\over g})d^3x \eeq Here  \beq \nabla_{\mu}
Z=\pdr_{\mu}Z+ZA_{\mu} \eeq $A_{\mu}$ being a gauge field valued in $U(n)$. The
WZ  term can replaced  by a Chern--Simons term for $A$. (This will introduce
some massive vector mesons into the theory, but at low energies it will be
equivalent to the earlier theory). \beq S(Z,A)=\half \int \tr
\nabla^{\mu}Z^{\dag}\nabla_{\mu}Z d^3x+ N\int \tr \sigma(Z^{\dag}Z-{1\over
g})d^3x+{k\over 4\pi}\int \tr [AdA+{2\over 3}A^3] \eeq This theory admits a
${1\over N}$ expansion that is renormalizable, a straightforward generalization
of Ref. \[arefeva]. More precisely, it has a non--trivial Ultra--Violet stable
fixed point. It is possible that the Chern--Simons term will lead to the
existence of a non--trivial fixed point as well.

This work is supported in part by the US Department of Energy Contract
No. DE-AC02-76ER13065.

\vfill\eject

{\it References}

\refis{II} G. Ferretti, S.G. Rajeev and Z. Yang, {\it Baryons as Solitons in
Three Dimensional Quantum Chromodynamics}, U. of R. preprint, April 1992.

\refis{balreview} A.P. Balachandran, TASI lectures, eds. F. Gursey and M.
Bowick (World Scientific, Singapore, 1985).

\refis{anderson} G. Baskaran and P.W. Anderson, Phys. Rev. B37 (1988) 580; P.B.
Wiegmann, Phys. Rev. Lett. 60 (1988) 821.

\refis{fradkin} E. Fradkin, Field Theories of Condensed Matter Systems
(Addison-Wesley, New York, 1991).

\refis{affleck} I.K. Affleck and J.B. Marston, Phys. Rev. B37 (1988) 3774.

\refis{wiegmann} P.B. Wiegmann, Phys. Scripta T27 (1989) 160.

\refis{andersen} P.W. Anderson, Science 235 (1987) 1196.

\refis{witten} E. Witten, Nucl. Phys. B160 (1979) 57.

\refis{forthcoming}S.G.Rajeev, { Two Dimensional Meson Theory}, U. of Rochester
Preprint UR-1252 (1992).

\refis{vafawitten} C. Vafa and E. Witten, Phys. Rev. Lett. 53 (1984) 535; Nucl.
Phys. B234 (1984) 173.

\refis{wittcurralg} E. Witten, Nucl. Phys. B223 (1983) 422.

\refis{lee} L. Brekke and T. Imbo, Harvard preprint HUTP-92/A020.

\refis{djt} S. Deser, R. Jackiw, S. Templeton, Phys. Rev. Lett. 48 (1982) 975;
Ann. Phys. 140 (1982) 372.

\refis{anr} A.N. Redlich, Phys. Rev. Lett. 52 (1984) 18; Phys. Rev. D29 (1984)
2366.

\refis{noren} S. Coleman and B. Hill, Phys. Lett. 159B (1985) 184.

\refis{thooft} G. 't Hooft, Proceedings of 1979  Cargese  School, eds., G. 't
Hooft (Plenum, New York, 1980).

\refis{colewitten} S. Coleman and E. Witten, Phys. Rev. Lett. 45 (1980) 100.

\refis{hagen} C.R. Hagen, Univ. of Rochester preprint UR-1212 (1991).

\refis{witten2d} E. Witten, Comm. Math. Phys. 92 (1984) 455.

\refis{bhattraj} G. Bhattacharya and S. G. Rajeev, Nucl. Phys. B246, 157
(1984).

\refis{milnor} J.W. Milnor and J.D. Stasheff, Characteristic Classes (Princeton
Univ. Press, Princeton, 1974).

\refis{arefeva} I.Y. Arefeva, Theor. Math. Phys. 31 (1977) 279.

\bye